\documentclass[prl,aps,showpacs,reprint,superscriptaddress,byrevtex,amsfonts,amssymb,amsmath]{revtex4-1}
\usepackage{bm,float}
\usepackage{amsxtra,amstext,latexsym}
\usepackage{times}
\usepackage{mathptm}
\usepackage{wasysym}
\usepackage{xcolor}
\usepackage{graphicx}
\graphicspath{{pdf_box/}}
\begin{document}

\title{Direct observation of current in type I ELM filaments on ASDEX Upgrade}
\author{N.~Vianello}
\affiliation{Consorzio RFX, Associazione Euratom-ENEA sulla Fusione, Padova, Italy}
\author{V.~Naulin}
\affiliation{Association EURATOM-Ris{\o} DTU,
OPL-128 Ris{\o}, DK-4000 Roskilde, Denmark}
\author{R.~Schrittwieser}
\affiliation{Association EURATOM/\"OAW, Institute for Ion Physics and Applied Physics, University of Innsbruck, Austria}
\author{H.W.~M\"uller}
\affiliation{Max-Planck-Institut f\"ur Plasmaphysik, EURATOM Association, Garching, Germany}
\author{M.~Zuin}
\affiliation{Consorzio RFX, Associazione Euratom-ENEA sulla Fusione, Padova, Italy}
\author{C.~Ionita}
\affiliation{Association EURATOM/\"OAW, Institute for Ion Physics and Applied Physics, University of Innsbruck, Austria}
\author{J.J.~Rasmussen}
\affiliation{Association EURATOM-Ris{\o} DTU,
OPL-128 Ris{\o}, DK-4000 Roskilde, Denmark}
\author{F.~Mehlmann}
\affiliation{Association EURATOM/\"OAW, Institute for Ion Physics and Applied Physics, University of Innsbruck, Austria}
\author{V.~Rohde }
\affiliation{Max-Planck-Institut f\"ur Plasmaphysik, EURATOM Association, Garching, Germany}
\author{R.~Cavazzana}
\affiliation{Consorzio RFX, Associazione Euratom-ENEA sulla Fusione, Padova, Italy}
\author{M.~Maraschek }
\affiliation{Max-Planck-Institut f\"ur Plasmaphysik, EURATOM Association, Garching, Germany}

\author{the ASDEX Upgrade team}
\affiliation{Max-Planck-Institut f\"ur Plasmaphysik, EURATOM Association, Garching, Germany}

\date{\today}

\begin{abstract}
Magnetically confined plasmas are often subject to relaxation oscillations accompanied by large transport 
events. This is particularly the case for the high confinement regime of tokamaks where these events are termed 
edge localized modes (ELMs). They result in the temporary breakdown of the high confinement and lead to high 
power loads on plasma facing components. Present theories of ELM generation rely on a combined 
effect of edge current and the edge pressure gradients which 
result in intermediate mode number 
($n \cong 10-15$) structures (\emph{filaments})  
localized  in the perpendicular plane 
and extended along the field line. It is shown here by 
detailed localized measurements of the magnetic field perturbation associated to an 
individual type I ELM filament that these filaments carry a substantial current. 
\end{abstract}

\maketitle

Edge localised modes (ELMs) are short (ms) breakdowns of the high confinement regime (H-mode), 
which is envisaged to be used in power producing tokamaks. Due to the high power fluxes associated, 
the presence of ELMs poses demands on the design of plasma facing components, that are hard to meet. Thus 
understanding, with the aim of controlling, ELM events is one of the foremost 
priorities in fusion research.
Moreover, apart from the interest for fusion oriented plasmas, the interest in ELM physics 
is enhanced by some fascinating analogies with sporadic explosive events as observed for example in solar flares \cite{Fundamenski:2007p1734} or 
in magnetic substorms \cite{Lui:2000p4196}.

ELMs are at present thought to originate from a combination of current 
and pressure gradient driven MHD modes \cite{Snyder:2005p555}, and are  
known to result in a medium number ($n\approx 10 - 15$) of structures 
\cite{Kirk:2006p2233,Fenstermacher:2005p3512, Herrmann:2007p211,Kirk:2006p125}, localised in the perpendicular plane and extended 
along the magnetic field. 
These filaments travel through the Scrape Off Layer (SOL) and have been measured using 
Langmuir probes on various machines, see e.g. \cite{Leonard:2006p1221}, and observed 
through high speed cameras \cite{Kirk:2006p125} or Gas Puff Imaging diagnostic \cite{Maqueda:2009p4189}.
In this letter we present results of direct measurements of all three components of 
the magnetic field perturbations associated to ELM filament structures
in the SOL together with an estimate 
of the current carried by filaments.
 
Magnetic fluctuations associated with ELMs are usually believed to originate from MHD activity. 
Measurements of the magnetic activity have previously been performed using magnetic 
pickup coils close to the vessel wall \cite{Kirk:2006p125,Takahashi:2008p3093} or on insertable 
probes \cite{Herrmann:2008p4203}. 
These measurements generally take place far from the filaments in comparison to their radial extent. 
This makes it difficult to observe the magnetic perturbation going along with individual filaments 
and to examine the magnetic fine structure of the ELMs. Both would result in important information 
such as the excursion 
of magnetic field lines from their equilibrium position and if the ELMs are 
associated with reconnection events \cite{Fundamenski:2007p1734} . 

\begin{figure}[!t]
\centering
\includegraphics[width=\columnwidth]{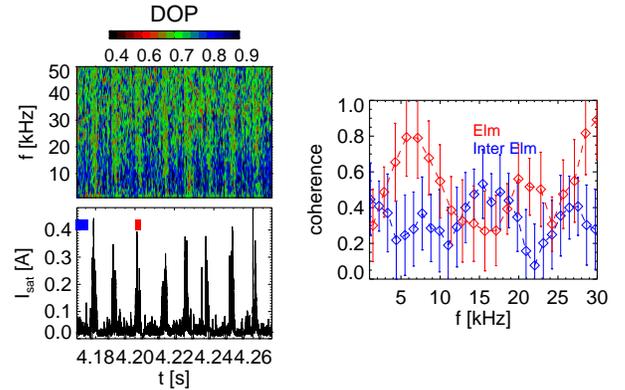}
\caption{Top: Degree of Polarization (DOP) analysis as a function of time and frequency.  Bottom:
Ion saturation current (I$_{sat}$). Right: Coherence between I$_{sat}$ and $b_p$ computed in 
the ELM and inter-ELM phases. The time intervals are shown in the bottom panel with colored boxes. } 
\label{fig:dop}
\end{figure}
Here we report on measurements of type I ELMs carried out at ASDEX Upgrade (AUG) tokamak by means 
of a newly constructed probe head described in detail elsewhere
\cite{Ionita:2009p4410}. The probe head 
consists of a cylindrical graphite case 
(\diameter=60 mm) which holds six graphite pins.  One of the tips  
measured the ion saturation current, one tip was swept, whereas all the others had been kept floating. 
Inside the case, 20 mm behind the front side, a magnetic sensor 
measuring the time 
derivative of the three components of the magnetic field is mounted. The sensor has a measured bandwidth of 1 MHz. 
A similar probe with combined electrostatic and magnetic signals has
already been used in AUG \cite{Herrmann:2008p4203}, but with a wider 
magnetic coil and a larger radial separation between electrostatic and
magnetic sensors.
The probe head, mounted on the fast 
reciprocating midplane manipulator, is inserted from the low field
side of the torus for 100 ms 12 mm inside the limiter position. 
The data have been obtained in type-I ELMy plasma discharges
(\# 23158,23159,23160,23161,23163) with a toroidal magnetic field of -2.5 T, 0.8 MA 
of plasma current, 6.5 $\times10^{19}$ m$^{-3}$ 
central electron density and  a $q_{95}$ value of 5.2.

During the analysis we have used the ion saturation 
current to infer the passing of a type-I ELM 
filament structure in front of the probe, as done for example in
\cite{Endler:2005p335}.
\begin{figure}[!t]
\centering
\includegraphics[width=\columnwidth]{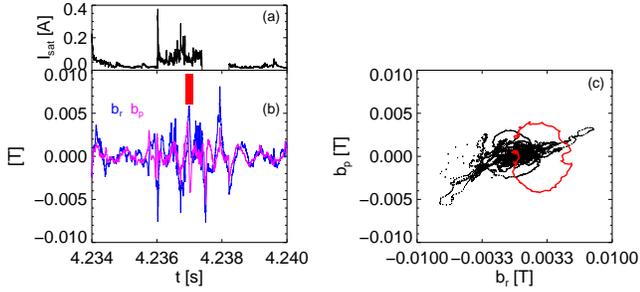}
\caption{(a)Ion saturation current (b) Poloidal and radial component
  of the magnetic field. The red box highlights the time interval when
  the two components change their phase relation. (c) Hodogram of the radial/poloidal magnetic field component. 
The closed loop in red refers to the time intervals highlighted in panel (a,b).} 
\label{fig:two}
\end{figure}
In analyzing the ELM events we 
employ the idea that the magnetic signal during the ELM can be 
separated into different frequency components. 
Higher frequencies, above few hundred kHz, we presume to be generated mostly by Alfv{\'e}nic activity or high frequency turbulence. 
These are not considered in the present analysis, also because of the 
frequency cut-off by the graphite shielding \cite{Ionita:2009p4410}. 
Additional MHD activity will still be present in the signal at frequencies below 20 kHz, 
but during ELM filaments most of the signal is presumed to originate from 
slowly varying currents convected with the filaments. This is 
justified by the so-called Degree of Polarization (DOP)
analysis \cite{Samson:1980p3485,*Santolik:2003p3492}, which  
is a test for a plane wave ansatz, quantifying how well the relation $\mathbf{k}\cdot\mathbf{B} = 0$ is satisfied.
It is based on the 
evaluation and diagonalization 
of the spectral matrix $S=\langle B_{i}^{*}B_{j}\rangle$, calculated in Fourier space. 
The DOP represents 
a measure to determine whether $S$ represents a pure state quantifying how much one single eigenvector approximates the state. 
A high value of DOP implies that the 
fluctuations considered are coherent over several wavelengths, and thus  
the method can be used to distinguish between propagating modes and coherent localized 
fluctuations. Figure 
\ref{fig:dop}, panel (a), shows the results of DOP analysis as a function 
of frequency and time. The temporal evolution of 
the ion saturation current is depicted in the lower panel. 
A sudden drop in the DOP is observed
in correspondance with a steep increase of $I_{sat}$ signals: 
this implies that the magnetic field fluctuations during an ELM can 
be better represented by coherent structures than by plane
wavepackets. 
\begin{figure}[!t]
\centering
\includegraphics[width=\columnwidth]{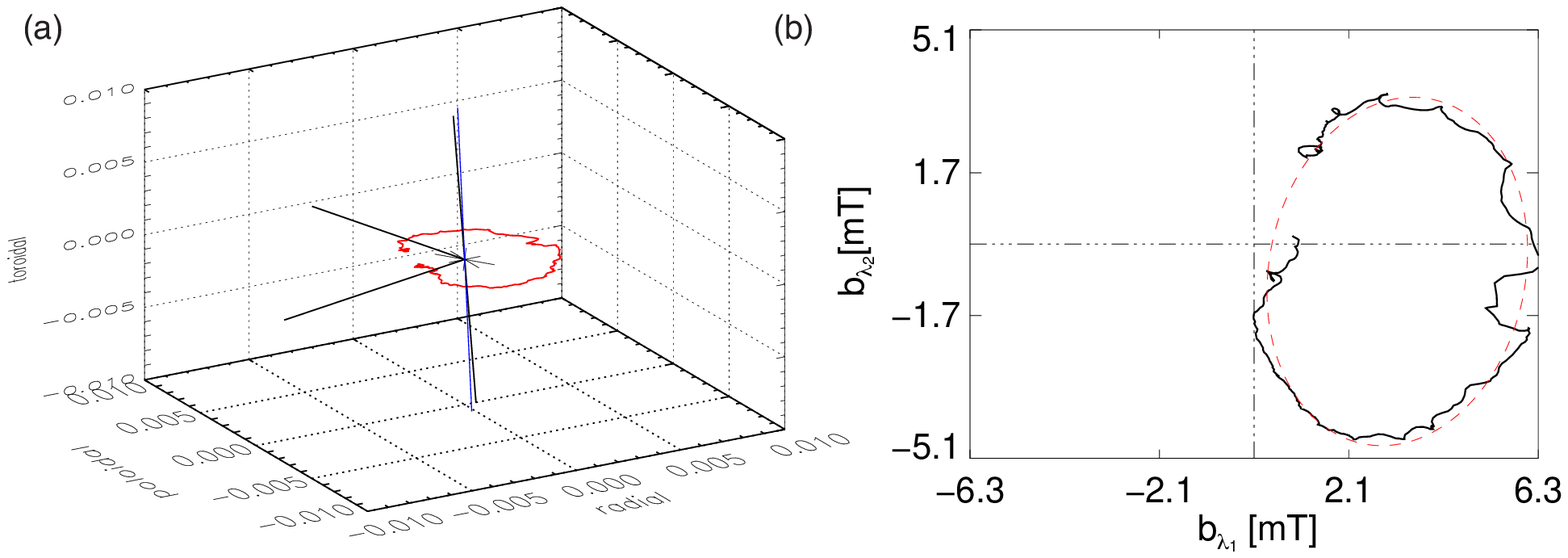}
\caption{(a)Trajectory of ELM filament associated magnetic field excursions in all three spatial directions. The direction 
of minimum variance is shown (black line) together with the direction of the equilibrium field (blue line).
(b) Hodogram of the magnetic perturbation associated with ELM current filament reconstructed in the maximum variance plane. The red line shows the elliptical fit.}
\label{fig:fig3}
\end{figure}
A coherence analysis between the ion saturation current and the
poloidal component of the magnetic field (Fig. \ref{fig:dop} (c)) reveals an increase of coherence between the two signals during the ELM activity.
A closer 
look on the behavior of magnetic fluctuations associated to an ELM
event is shown in Figure 
\ref{fig:two} (b). 
When the magnetic fluctuation amplitude increases, the radial and
poloidal components 
($b_{r}$ and $b_{p}$ ) change their phase 
relation as highlighted by the color box. 
We now consider the hodogram of the perpendicular
magnetic perturbation 
(Fig. \ref{fig:two} (c)), 
 i.e. the magnetic field perturbation trajectory in the $b_{r}-b_{p}$ plane, during the passing 
of the ELM. Here one closed orbit corresponds to the time interval marked in Figure \ref{fig:two} b. 
Closed loops are compatible 
with the passing of current filaments in front of the probe. 
Outside the ELM, 
the magnetic field perturbation 
exhibits an almost linear polarization in the perpendicular plane.
This shows clearly that magnetic activity in between ELMs  (wavelike) differs qualitatively from the magnetic 
perturbation during an ELM, which seems to be due to the motion of current filaments. 

The measurements of all three components of the magnetic perturbation  
allows to check the alignment of the current filaments directly. In Figure \ref{fig:fig3} (a) 
the trajectory of the magnetic field excursion during an ELM event in all three magnetic field components is shown. The time 
interval selected corresponds to the one highlighted in Figure 
\ref{fig:two}. A closed 
elliptical loop lying  in a plane slightly tilted with respect to the
local frame of reference is observed, as expected for filamentary structures. The 
direction normal to this plane can be determined by using the 
minimum variance analysis \cite{Weimer:2003p2577}. 
The method is based on the solution of the eigenvalue problem $\sum_{\nu=1}^{3} M_{\mu\nu}^{B}n_{\nu} = \lambda n_{\mu}$
where $M_{\mu\nu}^{B}=\langle B_{\mu}B_{\nu}\rangle -\langle B_{\mu}\rangle\langle B_{\nu}\rangle$ is 
the \emph{Magnetic Variance Matrix}, the brackets indicating the mean 
values averaged over the time the structure spends traveling in front of the probe, 
$\mu,\nu$ = 1,2,3 denoting the cartesian components of the 
$X,Y$ and $Z$ of the chosen system, $\lambda$ and $n$
  being respectively 
the eigenvalues and 
eigenvectors of the system. 
\begin{figure}[!h]
\centering
\includegraphics[width=.55\columnwidth]{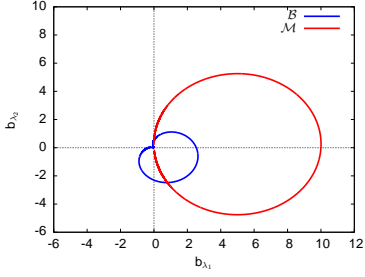}
\caption{Hodograms calculated for a monopolar current distribution ($\mathcal{M}$) and 
bipolar one ($\mathcal{B}$).}
\label{fig:hodomodel}
\end{figure}  
The eigenvectors corresponding to the eigenvalues 
$\lambda_{1}$, $\lambda_{2}$ and $\lambda_{3}$, represent  the direction of maximum, 
intermediate and minimum variance of the magnetic field respectively. The eigenvector corresponding 
to the minimum eigenvalue corresponds to the direction of minimum variance and is 
normal to the plane spanned by the magnetic field perturbations. This direction is observed
to be parallel to that of the equilibrium magnetic field, shown 
in the same plot by a blue line. 
This confirms the hypothesis  that the ELM filament is aligned with 
the equilibrium magnetic field. 
From the sense of the polarization the direction 
of the current is found to be colinear with the plasma current. 
The method allows the determination of the rotation 
matrix, so that the hodogram in the plane perpendicular 
to the current filaments may be reconstructed. This hodogram is shown
in Figure \ref{fig:fig3} (b) together with a fit to an ellipse. 
The shape of the hodogram calculated in the rotated frame of reference can be used to determine the type of current 
distribution associated to an ELM. 
Theories have both proposed ELM filaments to be associated to monopolar  
\cite{Kirk:2006p2233,Myra:2007p2641} and bipolar current distributions \cite{Rozhansky:2008p3518}. 
Up to now no clear 
evidence of one mechanism predominant with respect to the other has been reported. 
Figure \ref{fig:hodomodel} shows the anticipated shape of the hodogram in the bipolar and monopolar cases. 
The hodogram in the bipolar case exhibits a cardioid-like shape with 
a cusp at the origin, which represents a distinct signature and which can not be recognized in the experimental 
data (see Fig \ref{fig:fig3} (b)). To reinforce the statement we note that a bipolar-like hodogram with 
the presence of a cusp has been previously recognized in \cite{Spolaore:2009p4115} (cfr. fig 3(b)) where it has been 
associated to a direct measurement of a bipolar current structures. We emphasize, however, that the filaments 
observed in \cite{Spolaore:2009p4115}  are of a completely different
origin 
as induced by Drift-Alfv\'en turbulence \cite{Vianello:2010p4670}. A further corroboration of the hypothesis of 
monopolar current filaments results from considering the quadrants
occupied by the magnetic field trajectory. 
Indeed while the monopolar current  hodogram regularly occupies two of the quadrants, the bipolar one always spans three of the quadrants: this latter behavior is not observed in 
the experimental data as can be verified by comparing Figure \ref{fig:fig3} (b) and Figure \ref{fig:hodomodel}.
These results make us confident that the magnetic fluctuations observed are indeed generated by 
a monopolar current distribution. Under this basic assumption 
the current carried by this filament may be estimated.
 \begin{figure}[!h]
\centering
\includegraphics[width=.9\columnwidth]{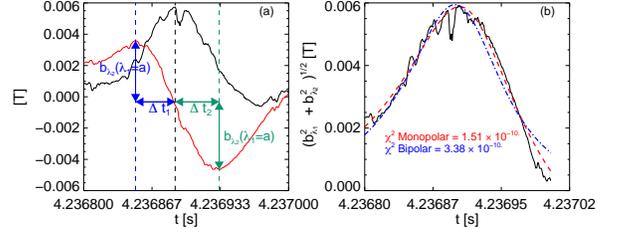}
\caption{(a) Time traces of the two perpendicular components of magnetic field during an ELM filament. 
(b) Total perpendicular magnetic field as a function of time. Superimposed 
the results of a fit for an expected monopolar 
(red) and bipolar (blue) current distribution both modulated by an
exponential decay. }
\label{fig:fig5}
\end{figure}
In the rotated frame of reference for a circular monopolar symmetric filament 
drifting in the $\lambda_{1}$ direction, the two perpendicular magnetic field components may be 
written as
$b_{\lambda_{1}}=\frac{-r_{0}B_{0}a}{\lambda_{1}^{2}+a^{2}}$ and  $b_{\lambda_{2}}=-\frac{r_{0}B_{0}\lambda_{1}}{\lambda_{1}^{2}+a^{2}}$
where $B_{0}= \frac{\mu_{0}I_{0}}{2\pi r_{0}}$ and  
$a$ is the distance between the trajectory of the center of the
filaments and the probe, representing also the distance of closest approach, 
$r_{0}$ is the radius of the filaments and $I_{0}$ its current. 
The distance $a$ can be approximated, assuming the filament to propagate with a 
constant velocity along the $\lambda_{1}$ direction: $a = \Delta t v_{\lambda_{1}}$ where 
$\Delta t$ is the time delay between the maximum of $b_{\lambda_{1}}$
and the 
maximum/minimum of $b_{\lambda_{2}}$, 
where $b_{\lambda_{1}}$ and $b_{\lambda_{2}}$ are the projections of the magnetic field in the maximum-intermediate variance plane. 
Thus within these approximations the current may be estimated noting
that $|b_{\lambda_{2}}(\lambda_{1}=a)| = \frac{1}{2}\frac{\mu_{0}I_{0}}{2\pi a} = \frac{\mu_{0}I_{0}}{4\pi \Delta t v_{\lambda_{1}}}$
Applied to 
the experimental data, this computation is equivalent to calculate 
the quantities depicted in Figure \ref{fig:fig5}, where $\Delta t$ has been 
calculated both at the maximum and at the minimum of $b_{\lambda_{2}}$. The two values $\Delta t_{1}$ and 
$\Delta t_{2}$ are equal 
to 38 $\mu$s and 42 $\mu$s respectively.
The estimate of the current relies on the knowledge of local
  velocity in the $\lambda_1$ direction. This propagation has been
  reported for ASDEX Upgrade, e.g.
  \cite{Herrmann:2008p4203,kirk:vr}. The 
  experimental setup does not allow a reliable local measurements of
  $v_r$ or $v_p$, neither can we rely in the presented shots on correlation analysis of wall
  mounted probes (which are 160 degrees separated from the
  manipulator in the toroidal direction). Most likely these signals
  are indeed dominated by MHD mode activity in the confined
  plasma rather than current filaments in the SOL. As
the best approximation we thus assume as radial propagation the most
probable value $v_r = 1.2$ km/s as determined in \cite{kirk:vr} and
calculate $v_{\lambda_1} = v_r/\cos(\angle(\lambda_1 - r))$, where
$\angle(\lambda_1-r)$ is the nagle between $\lambda_1$ and the radial
direction which for the present case is of the order of 7$^{\circ}$. From
\cite{kirk:vr} we also estimate a standard
deviation of $v_r$ as $\sigma_{v_r} = 700$ m/s, which ensure that 71 \% of the events are
observed within $(v_r \pm \sigma_{v_r})$. 
With these values
the average distance of closest approach is approximately 4 cm and
using the average value between the minimum and maximum of
$b_{\lambda_2}$ we obtain an estimate of 1.9 kA.
\begin{figure}[!t]
\centering
\includegraphics[width=.85\columnwidth]{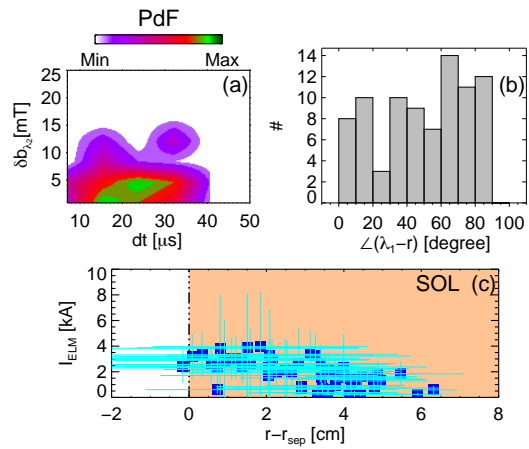}
\caption{(a) Joint PDF of $\Delta t$ and $b_{\lambda_2}(\lambda_1 =
  a)$ (b) )Histogram of the angle between $\lambda_1$ and the radial direction.
(c) Filaments current estimate \textit{vs} distance from the separatrix estimated
with $v_r = 1.2$ km/s. Error
bars results from the propagation of velocity uncertainty in the
estimate of $a$ and $I$.}
\label{fig:fig6}
\end{figure} 
The total perpendicular field $\sqrt{b_{\lambda_{1}}^{2}+b_{\lambda_{2}}^{2}}$ is shown 
in figure \ref{fig:fig5} (b). The magnetic field variation can be fitted with a function
of the form
$f = \left(\frac{\alpha}{\sqrt{(t-t_{0})^{2}+\beta}} -
    \gamma\right) e^{-(t-t_{0})/\tau_{e}}$, $t_{0}$  being the time instant 
corresponding to the maximum of the total perpendicular field and
$\tau_{e}$ the e-folding time, and assuming the magnetic field is the
response from a passing monopolar current filament. The good quality of the fit provides additional 
support for the monopolar nature of the filament, also 
comparing with the fit expected from a bipolar current
distribution shown in blue line which exhibits an higher $\chi^2$.
The average decay time is determined as $\tau_{e}\approx 200$ $\mu$s.
In order to increase the statistic reliability of the previous
estimate we have analyzed events from 5 different  
shots with similar conditions. The results are shown in Figure
\ref{fig:fig6}. In the panel (a) the joint PDF of the independent
experimental values $\Delta t$ and $b_{\lambda_2}$ used for the
current evaluation is shown, highlighting how the bulk of the
filaments have values of approximately 25 $\mu s$ and 5 mT. In the
panel (b) the histogram of the angle between $\lambda_1$ and 
the radial direction is shown, showing that the $\lambda_1$ direction
is a complex combination of radial and poloidal propagation. Finally
in the panel (c) the current estimated accordingly to the
aforementioned formula is plotted \textit{vs} the position of
the filaments with respect to the separatrix, taking into account the position of the manipulator and
the distance of closest approach estimated. The errors shown represent
the influence of the velocity uncertainty on the estimate of $a$ and
consequently on $I$. The bulk of the distribution of these filaments
are thus observed in the SOL, even taking into account the
uncertainty on $v_r$: the median of the distribution of the filaments
detected in the SOL within the error in their position is 1.4
kA, corresponding to a $j_{\parallel} \approx 4.5 MA/$m$^2$ for 1 cm
radius filaments.
These values are consistent with measurements of edge current as seen for example in 
\cite{Thomas:2005p4575}. It has already been supposed in
\cite{wolfrum} that, 
in case ELM breakdown is driven by peeling instabilities, 
this will lead to a 
flattening of the current density profile over the separatrix and a
subsequent 
increase of currents flowing into the SOL. These 
currents will be nearly force-free and will be accompanied by
poloidal 
halo currents closing through the divertor tiles. 
Actually these currents have been measured
\cite{Ionita:2009p4410,mccarthy} with 
values up to few tens of kA, supporting 
the presence of rapid flow of toroidal currents  from the plasma into the SOL
The resulting 
histogram may also be compared with the estimate for ELM filaments in JET 
given in \cite{migliucci}: in this case the most probable current has been found of the order 
of 450 A, but with a lower radial velocity postulated. 
For completeness it must be noted that the same current density  was estimated in \cite{Herrmann:2008p4203}
assuming the magnetic perturbation 
to be induced by a rotating helical structure with a bi-directional current close to, but still inside, the separatrix.
The  value found is higher than the measured $j_{sat}$ current density to the ion-biased tip (approximately 50 kA/m$^{2}$).
Concluding in this letter we have provided evidence that ELM filaments
carry considerable currents 
for which 
we have found a reasonable estimate. 
The magnetic signals during ELM filaments differ substantially from wave activity 
in between ELMs. We have shown that the current in the filaments is 
co-aligned to the plasma current and approximately of a magnitude as
expected for the edge. 
The current flows along 
the unperturbed magnetic field lines and has a unidirectional
nature. This poses the question 
where the filament currents 
close in the SOL and why such high current densities are sustained in the ELM filaments.
We hope that future experiments will contribute to answer these
questions, which will throw 
new light on the instability mechanisms for ELMs.\\
This work, supported by the European Communities under the contract of
Associations between 
EURATOM and ENEA, Ris\o/DTU, \"OAW Innsbruck, and IPP
Garching, 
was carried out within the framework of the European Fusion
Development Agreement. 
This work was also supported by project P19901 of the Austrian Science Fund (FWF).

%

\end{document}